\documentstyle[prl,twocolumn,aps,epsfig]{revtex}

\begin{document}

\title{Subband population in a single-wall carbon nanotube diode}

\author{R. D. Antonov and A.T. Johnson } 

\address{Department of Physics and Astronomy, University
of  Pennsylvania, 209 S. 33rd St., Philadelphia, PA  19104,
USA}

\date{\today}
\maketitle

\begin{abstract}
We observe current rectification in a molecular diode consisting of a semiconducting single-wall carbon nanotube and an impurity. One half of the nanotube has no impurity, and it has a current-voltage (I-V) charcteristic of a typical semiconducting nanotube. The other half of the nanotube has the impurity on it, and its I-V characteristic is that of a diode.  Current in the nanotube diode is carried by holes transported through the molecule's one-dimensional subbands. At 77 Kelvin we observe a step-wise increase in the current through the diode as a function of gate voltage, showing that we can control the number of occupied one-dimensional subbands through electrostatic doping.        
\end{abstract}
 

%73.50.-h (elec trans in low-d struc)
%73.23.-b (elec struc of mesosystems)
%71.20.Tx,  73.40.Sx (elec props of metal-semi-metal strucs)
%73.61.Wp (elec struc of full & related maters)

Electronic circuit elements, such as resistors, transistors and diodes, made with carbon nanotubes could some day become components in molecular scale electronics.  Single-wall carbon nanotubes (SWNTs) are especially suitable for use because they can be either semiconducting or metallic depending on the geometry of the carbon atoms on the circumference of the tube \cite{nhamada,nsaito,nmintmire,nwildoer,odom_nature}.  Additionally, atomic scale defects, molecular adsorbates, mechanical deformations, and dopants are predicted to affect a tube's electronic properties \cite{crespi_junctions,kane1,kawazoeapl,clauss,venema_sci} and could be used to tailor device operation. Recent experiments have demonstrated nanotube-based electronic devices, such as quantum wires and single electron transistors at low temperature \cite{nsander1,nbockrath} and room temperature field effect transistors (FETs) \cite{nsander2,martel_apl}.  Scanning tunneling microscope (STM) measurements on bulk nanotube ``mats" suggest that defects can {\it locally} alter the electronic properties of a nanotube \cite{phil}. STM experiments also provide compelling evidence that impurities adsorbed onto a tube can have a similar effect by acting as strong electron scatterers \cite{clauss}. Here we report current rectification in a circuit consisting of a single semiconducting nanotube with an impurity present on one part of the tube.  In contrast, a segment of the {\it same} tube without the impurity exhibits symmetric current-voltage (I-V) characteristics.  We argue that rectification results from the local effect of the impurity on the tube's band structure \cite{kawazoeapl}.  By varying an external back-gate voltage, we can tune the number of current carrying, one-dimensional electronic subbands in the nanotube diode by ``electrostatic doping''.
  
\begin{figure}
\centerline{\epsfig{figure=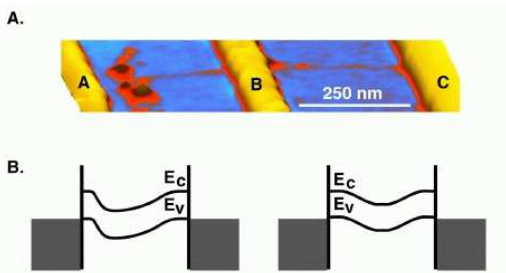,width=7cm}}
\caption{{\bf A.} AFM image of a $1$ nm high, single-wall carbon nanotube connecting three gold electrodes: A, B and C. {\bf B.} Left: Band structure for the tube segment with impurity (electrode pair AB), with zero bias voltage. The impurity acts like an n-type dopant that locally lowers the energy of the bands. Right: Band bending in the clean nanotube segment (BC) with zero bias voltage.}
\label{sample_pic}
\end{figure}

We electrically contact single nanotubes circuits using a method similar to that of Ref.\ \cite{nsander1}.  An Atomic Force Microscope (AFM) image of the sample (Fig.\ \ref{sample_pic}A) shows a $1.0 \pm 0.1$ nm high SWNT contacting three gold electrodes A, B and C. We believe the nanotube has a larger diameter, but the height as measured by AFM is reduced due to the force between tip and sample. The tube segment between electrodes B and C is clean, but there is a visible impurity deposit near electrode A \cite{note1}. The circuit was patterned on a heavily doped Si substrate covered with a $100$ nm insulating oxide layer, so the substrate acts as a capacitively-coupled gate electrode.  A solution of nanotubes in dichloroethane was deposited on the sample and pairs of leads checked for electrical contact.  Typically we measure resistance values between $0.5$ and $10$ M$\Omega$ when a single tube or tube bundle contacts two electrodes, where the two resistance ranges correspond to metallic and semiconducting tubes, respectively. The room temperature contact resistance, estimated from samples where a tube contacts four electrodes, is about $0.5$ M$\Omega$.

\begin{figure}
\centerline{\epsfig{figure=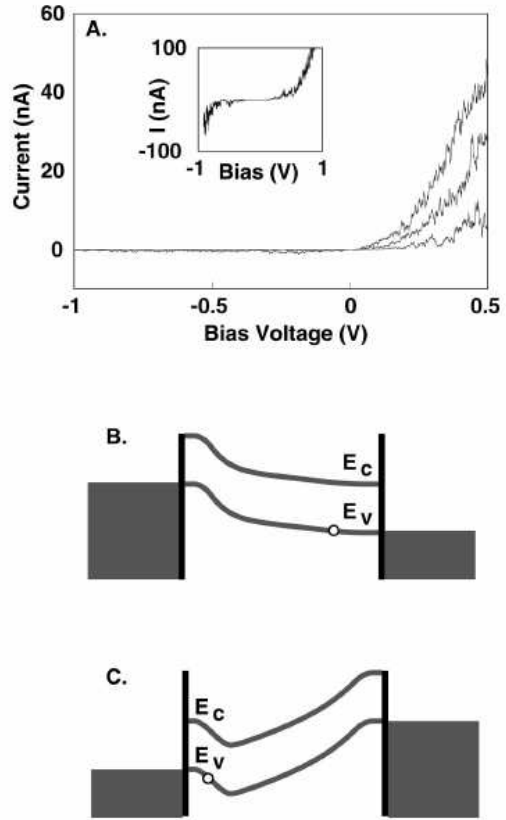, width=7cm}}
\caption{{\bf A.} Highly asymmetric room temperature I-V characteristics of the nanotube section with the impurity (AB).  The $V_g$ values are $-7, 0$ and $+7$ V (top to bottom).  Inset: Symmetric I-V curve for the clean tube segment (BC). {\bf B.} Proposed band diagram of the nanotube diode under forward bias.  Space charge build up near the left contact eliminates the effect of the impurity potential. The current in the diode is carried by holes injected from the positive electrode on the right. {\bf C.} Bands in the reverse-biased diode showing space charge build up on the right side of the device. Holes injected from the left contact cannot pass the impurity-induced barrier.}
\label{plots1}
\end{figure}

The inset to Fig.\ \ref{plots1}A is the room temperature I-V characteristic of the clean tube segment BC with zero voltage on the gate. The I-V curve is symmetric around the origin with a gap of about $0.5$ V \cite{nhamada,nsaito,nmintmire}, and the device conductance can be sharply increased by applying a negative gate voltage $V_g$ (not shown), consistent with what is expected for a semiconducting nanotube FET \cite{nsander2,martel_apl}. In marked contrast are the I-V curves for tube segment AB containing the impurity (Fig.\ \ref{plots1}A).  The I-Vs are highly asymmetric: for any $V_g$, current flows through the nanotube readily when lead A is biased negatively with respect to lead B (forward bias), but only for voltages larger than $1$ V under reverse bias.  At positive gate voltage, the current through the nanotube decreases, indicating that transport in the diode is dominated by holes.   Below we present a model where the impurity locally alters the  nanotube electronic structure, leading to the highly asymmetric I-V for tube segment AB. This set of experiments provides the strongest, most direct evidence to date supporting the key theoretical prediction that defects and chemical impurities can be used to create functional nanotube electronic circuits \cite{crespi_junctions,kane1,kawazoeapl}.  

Our measurements can be explained using a modified BARITT diode model, inspired by Tans's model for a nanotube FET \cite{nsander2}. We first review the model for the clean tube segment BC. At the contacts, the valence band of the nanotube is pinned to the Fermi level of the leads (Fig.\ \ref{sample_pic}C), possibly due to the mismatch of the nanotube and gold workfunctions. The energy bands bend to lower energy in the middle of the tube segment. This creates a potential barrier that is symmetric with respect to the two electrodes, ultimately leading to symmetric I-Vs. At low voltage bias, the electric field in the middle of the nanotube is zero, and charge flow through the tube is blocked.  For a sufficiently large bias, the electric field reaches through the tube \cite{sze,luryi_apl}, and holes injected at the positive electrode are transported to the drain.  This model predicts symmetric I-Vs with a gap, as observed for the clean tube segment BC.

We propose that the impurity on segment AB creates an n-type doped tube region, either by charge transfer or impurity-induced mechanical deformation, that localizes charge on the nanotube. The doped region distorts the valence band barrier, bringing the energy minimum close to contact A (Fig.\ \ref{sample_pic}B). This breaks the spatial symmetry of the device, and gives strongly rectifying I-V curves. When contact A is baised negatively with respect to B (Fig.\ \ref{plots1}B), space charge builds up near the left contact; the impurity barrier is suppressed, and current flows readily as  holes injected at B travel to contact A. When the bias is reversed, space charge builds up on the {\it right} side of the device, with little effect on the impurity barrier at A (Fig.\ \ref{plots1}C). Hole transport is blocked by the barrier, and the device remains insulating for this reverse-bias polarity. The presence of the impurity creates a nanotube diode with strongly asymmetric, rectifying I-Vs.

\begin{figure}
\centerline{\epsfig{figure=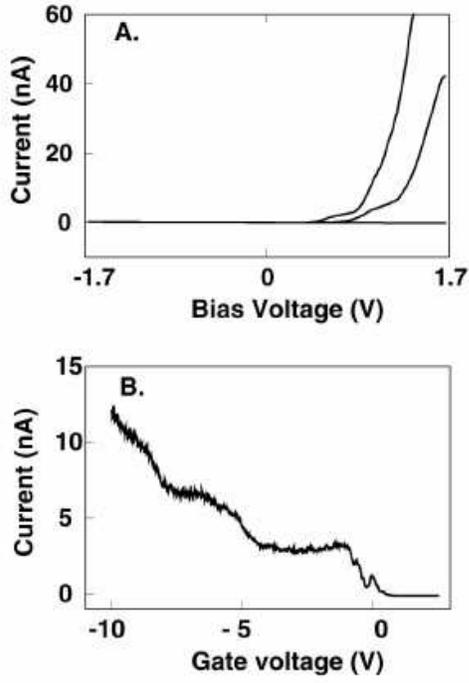, width=7cm}}
\caption{{\bf A.} I-V curves for the nanotube diode at 77K and $V_g = -5, 0, +5$ V (top to bottom).  Step-wise current increases with bias are due to sharp van Hove singularities in the density of states at the edge of 1D electronic subbands.  {\bf B.} Current vs. gate voltage in the nanotube at $1$ V forward bias.  Steps occur when another 1D subband can transport charge through the diode.}
\label{plots2}
\end{figure}

Current in all nanotube devices is thought to be carried by one-dimensional (1D) electronic subbands. We are able to clearly resolve the effect of these subbands on transport when the diode is cooled to 77 K. As shown in Fig.\ \ref{plots2}A, we observe well-defined current steps with increasing bias in the diode I-Vs, which we attribute to sharp van Hove singularities in the density of states (DOS) at the band edge energies of the 1D modes \cite{nhamada,nsaito,nmintmire}. The physics of this observation is essentially the same as earlier STM spectroscopy of clean tubes on gold substrates \cite{nwildoer,odom_nature}, but to our knowledge has not been seen in transport {\it along} tubes until now. Each step in the I-V corresponds to the availability of an additional subband for hole transport in the valence band. The $0.5$ V separation between current steps agrees with calculations of the subband separation in semiconducting nanotubes with diameter near $1.4$ nm.

The number of subbands participating in transport through the diode also varies with the gate voltage $V_g$, which induces charge on the nanotube and {\it electrostatically} dopes it. When the nanotube charge is varied by changing $V_g$, the energy of the tube Fermi level $E_F$ changes relative to the valence band edge. As $E_F$ approaches the edge of a 1D subband, the number of states available for transport increases sharply, leading to a distinct current step. We observe precisely such steps in the diode current versus gate voltage (I-$V_g$) curve measured at 77 K (see Fig.\ \ref{plots2}B). The current steps appear as linear bands in a color scale plot of the current as a function of $V_g$ and bias voltage (not shown), as expected for features associated with the nanotube DOS. A change of $0.1$ V in the bias voltage corresponds to a $1$ V change in $V_g$. This agrees with the ratio of the gate capacitance to the total nanotube capacitance (the capacitive ``lever arm'') of $0.1$ found for substrate gates in low-temperature Coulomb blockade experiments done on single-nanotube samples in our lab and elsewhere \cite{nsander1,nbockrath}. Similar steps in the I-$V_g$ curves are visible at room temperature, although superimposed on a larger noise level, consistent with the fact that the subband energy separation (about $0.5$ eV) is large compared to the thermal energy ($25$ meV). It is likely that at room temperature the transport signatures of the electonic subbands are masked by noise associated with charge traps on the Si substrate. 

In conclusion, we have made and characterized a molecular diode consisting of an individual semiconducting single-wall carbon nanotube and an impurity deposit.  Step-wise current modulation by an external gate voltage at 77 K shows that we can controllably populate the nanotube's one-dimensional, current-carrying subbands by electrostatic doping.  The nanotube diode also demonstrates that impurity defects can locally alter the nanotube band structure.  This suggests the possibility of building multiple devices on a single tube by controlled deposition of impurities.  In the future, local dopants might be placed or defects created at specific locations along nanotube circuits by scanning probe manipulation, providing the opportunity for molecular scale device engineering.     

The nanotube material used in this work was provided by the Smalley group at Rice University. We acknowledge useful discussions with Charles Kane, Gene Mele, Wilfried Clauss, Jacques Lefebvre, and James Hone. This work was supported by the NSF under award number DMR-9802560. A.T.J. recognizes the support of the Packard Foundation and an Alfred P. Sloan Research Fellowship.

\vspace{20mm}

\vspace{20mm}

\begin{references}
\bibitem{note1}
We are unable to determine the composition of the impurity that creates the diode. We make the circuit using as-grown nanotube material with significant impurity content, including particles of graphite, amorphous carbon, and carbon-encased metal catalyst.
\end{references}
\end{document}